\documentclass[11pt]{article}
\usepackage[margin=1in]{geometry}
\usepackage{graphicx}
\usepackage{booktabs}
\usepackage{amsmath}
\usepackage{hyperref}
\usepackage[numbers,sort&compress]{natbib}

\title{Optogenetic Modulation Fraction: A Robust Opsin-Presence Effect and an Unresolved Cell-Type Question, by Censoring-Aware Survival Analysis of Public Neuropixels Data}
\author{Siddhardha Nanda\\
Columbia University\\
\texttt{sn3199@columbia.edu}}
\date{November 2026}

\begin{document}

\maketitle

\begin{abstract}
The clearest result in this paper is simple: expressing channelrhodopsin-2 (ChR2) in an interneuron population substantially and robustly raises the fraction of locally recorded units whose spike timing is altered by light, relative to no-opsin controls, under every method and outcome definition we tried. Establishing that result rigorously, and discovering that two more eye-catching secondary claims did not survive the same rigor, is the actual content of this paper. We reanalyze eight public Allen Institute Visual Coding Neuropixels sessions (4{,}112 units), modeling the latency from light-pulse onset to first spike as a right-censored time-to-event outcome, testing it with a Cox proportional-hazards model (log-rank fallback) and a single Benjamini-Hochberg FDR correction pooled across all units, and assessing genotype dependence with a formal, small-cluster-robust statistical test (standard cluster-robust logistic regression plus an exact wild cluster bootstrap, since only eight session-level clusters are available). Because a Neuropixels probe records the whole local population, not only the genetically targeted cells, we report two complementary quantities: an unrestricted modulation fraction (any unit with an altered spike-timing hazard, direct or network-mediated) and a restricted direct-responder fraction (modulated units whose light-evoked latency is under 10 ms, an explicitly stated threshold, since no universal numeric standard exists in the optotagging literature). ChR2-expressing Sst/Pvalb sessions show a large, bootstrap-confirmed elevation over control under both metrics (unrestricted 0.23-0.37 versus 0.01-0.05 FDR-corrected, bootstrap p = 0.0312; restricted 0.077 versus 0.013 pooled, bootstrap p $<$ 0.0001). Two further claims, each initially suggested by a standard test, did not survive the exact small-cluster bootstrap and a direct formal check, respectively, and we report both retractions as results in their own right. First, Vip-ChR2's intermediate unrestricted fraction (approximately 0.06) appeared different from control and from Sst/Pvalb under cluster-robust logistic regression, but the bootstrap could not confirm either contrast (p = 0.125 and p = 0.0625), and restricting to direct responders resolves the Vip-versus-control question toward clear non-significance (bootstrap p = 0.56). Second, a descriptive pattern in the restricted data, that among modulated units Vip sessions had a higher proportion of putative direct responses (34\%) than Sst/Pvalb sessions (20-28\%), the opposite of the raw ordering, looked like a genuine cell-type finding until we applied the same cluster-robust bootstrap test used everywhere else in this paper to the proportion itself: none of the three genotype contrasts on this metric approach significance (p = 0.75-0.84), and pooled across sessions the true no-opsin control group has the \emph{highest} direct-response proportion of all three groups (42.6\%), exceeding Vip, because the same 10 ms criterion that misclassifies some network-mediated units also misclassifies background activity in controls, and because per-session rates vary enormously even within a single genotype (control sessions alone range from 0\% to 76\%). We report this as a cautionary result, not a confirmatory one: an apparent reversal that would have been a compelling headline evaporated under the identical small-cluster-robust method that confirmed the paper's actual central finding. Together with two earlier self-identified corrections (a composition-dependent artifact in a smaller session pool, and the standard-versus-exact-bootstrap discrepancy described above), this is now a four-correction paper, and we take that as the main methodological lesson: pooled estimates, standard cluster-robust inference, and descriptive secondary patterns in small-sample circuit-dissection studies all deserve the same formal scrutiny as a paper's headline claim, not less.
\end{abstract}

\section{Introduction}

Optogenetic circuit manipulation rests on an assumption that is almost never tested directly: that light delivered to a genetically targeted population reliably produces the intended effect. Altahini, Arnoux, and Stroh's review of open problems in the field names this directly, noting that most circuit-dissection studies report that a manipulation was applied without measuring what fraction of the targeted population actually responded~\citep{altahini2024optogenetics}. This is not an isolated complaint: direct comparative studies of microbial opsins have separately documented that spike-evoking fidelity varies with opsin choice, expression level, and intrinsic cell excitability, without converging on a standard metric for reporting it~\citep{mattis2011principles}, which suggests the quantification gap is a recognized, recurring problem rather than one reviewer's framing.

Time-to-event framing of neural spike timing is not itself new. Point-process and hazard-based models of spike trains, most notably the time-rescaling theorem of Brown and colleagues, have related spike timing to hazard functions for over two decades~\citep{brown2002timerescaling}. Closer to the present application, Luo and colleagues modeled light pulses and evoked spikes as two coupled point processes under a generalized linear model to predict spike timing from stimulation history~\citep{luo2016pointprocess}; that work addresses a prediction question (can future spikes be forecast from the input point process) rather than a quantification question (what fraction of a population responds at all, with a formal account of censoring and multiple comparisons), which is the gap we address here. Our contribution is therefore not the time-to-event framing itself, but its specific application to quantifying optogenetic modulation fraction with explicit right-censoring, a single pooled multiple-comparisons correction, and small-cluster-robust inference, none of which is standard in the optotagging literature's typical practice of reporting a median or quartile first-spike latency, or a Kolmogorov-Smirnov test against shuffled light-onset times, without a censoring model or a formal account of how many units were tested.

This question is also distinct from, and should not be confused with, a separate and more mature line of work on classifying cell type from extracellular recordings using opto-tagging as ground truth. NEMO uses multimodal contrastive learning on waveform and autocorrelogram features to classify cell type~\citep{nemo2025}, and Beau and colleagues report a deep-learning classifier reaching high accuracy on opto-tagged ground truth from waveform and inter-spike-interval statistics~\citep{beau2025cell}. Both are classification-accuracy contributions: given a unit, predict its cell type. Neither asks, nor is designed to answer, what fraction of a genetically targeted population responds to stimulation at all, which is the quantification-of-efficacy question this paper addresses.

We use the Allen Institute's publicly released Visual Coding Neuropixels dataset to ask a narrower, directly answerable version of this question: given a Cre-line and opsin combination already in wide circulation (Cre-driven ChR2), what fraction of recorded units shows a statistically defensible, censoring-aware response to light, and does this fraction depend on cell type rather than being approximately constant across interneuron classes?

One definitional point must be stated before any result, since it determines what "modulation fraction" actually measures. A Neuropixels probe records extracellularly from the whole local population near the electrode, not selectively from the genetically targeted, Cre-defined interneurons; Sst-, Pvalb-, and Vip-positive cells are each a minority of the units recorded in a given session. A unit's spike-timing hazard can therefore change after light onset either because that unit itself expresses ChR2 and is directly photo-activated, or because it receives synaptic input from a directly activated cell elsewhere in the local circuit, a network-mediated effect. Reporting a single "modulation fraction" without distinguishing these two cases conflates "what fraction of the targeted population responded," the question this paper's motivation names, with "what fraction of the whole local population was affected by stimulating the targeted population," a related but different quantity. We address this directly in Section~\ref{sec:direct} by restricting to a latency-based criterion for putative direct responses, rather than treating the unrestricted, population-level fraction as a stand-in for targeted-population response.

\section{Data}

We used eight sessions from the Allen Institute Visual Coding Neuropixels dataset, accessed via the AllenSDK \texttt{EcephysProjectCache} interface~\citep{siegle2021survey}. All eight are \texttt{functional\_connectivity} sessions, the session type that includes an optogenetic stimulation block. Sessions were selected by ascending recorded unit count (the eight smallest sessions of this type among the 26 available), which produced the following genotype composition: one Sst-IRES-Cre;Ai32 session (ChR2 in somatostatin-positive interneurons, 415 units), two Pvalb-IRES-Cre;Ai32 sessions (ChR2 in parvalbumin-positive interneurons, 450 and 529 units), two Vip-IRES-Cre;Ai32 sessions (ChR2 in vasoactive intestinal peptide-positive interneurons, 582 and 585 units), and three wild-type sessions with no Cre driver and no opsin expression (474, 531, and 546 units), serving as negative controls. This selection rule was fixed before the genotype composition was known; the resulting mix of cell types was not chosen to support any particular outcome. Pooled across all eight sessions, 4{,}112 units were tested.

Each session includes up to 300 light presentations (single pulses at 5 or 10 ms, 10 Hz pulse trains, raised-cosine ramps); one wild-type session recorded only 180 presentations using the same three stimulus types, which the analysis handles without modification.

\section{Methods}

\subsection{Survival framing}

\begin{figure}[htbp]
\centering
\includegraphics[width=0.6\textwidth]{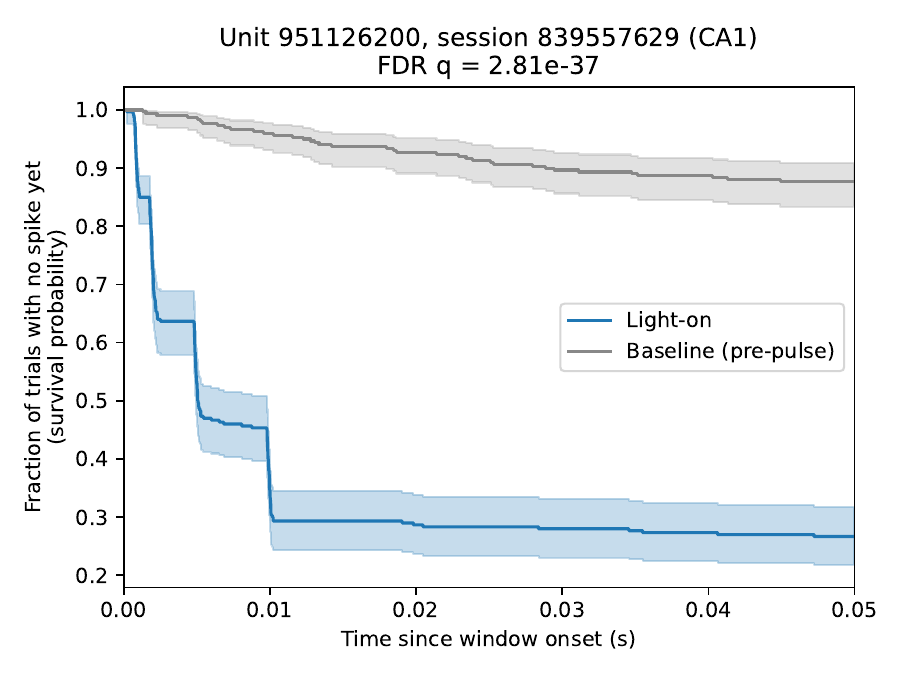}
\caption{Kaplan-Meier survival curves for a representative, strongly modulated unit, comparing the light-on window to the pre-pulse baseline window. Survival probability is the fraction of trials in which no spike has yet occurred.}
\label{fig:km_example}
\end{figure}

For each unit and each presentation, we measured the latency from pulse onset to the first spike within a fixed 50 ms post-pulse window. The window was chosen because the minimum inter-presentation interval across these sessions is 1.69 s, so the window never overlaps the next presentation's own baseline period, and because ChR2-driven interneuron responses are typically well under 20 ms, making 50 ms long enough that most true responses register as events rather than being censored by construction. If no spike occurred, the observation was right-censored at 50 ms rather than discarded or coded as a tied zero. A matched pre-pulse baseline window of the same duration immediately preceding each pulse onset served as the comparison condition. Figure~\ref{fig:km_example} illustrates this framing for a representative, strongly modulated unit: the light-on survival curve drops sharply within the first 10 ms as most trials produce a spike, while the baseline curve stays near 1.0.

\subsection{Per-unit significance testing and multiple-comparisons correction}

Each unit was tested with a Cox proportional-hazards model~\citep{cox1972regression} (group covariate: light versus baseline), implemented in \texttt{lifelines} 0.30.0~\citep{davidson2019lifelines}, falling back to the log-rank test~\citep{kaplan1958nonparametric} on non-convergence; for a single binary covariate this fallback is the asymptotic equivalent of the Cox score test, not a weaker substitute. Benjamini-Hochberg false discovery rate correction~\citep{benjamini1995controlling}, via \texttt{statsmodels} 0.15.0~\citep{seabold2010statsmodels}, was applied once, pooled across all 4{,}112 units, rather than separately within each session. Allen Institute data were accessed via AllenSDK 2.16.2; \texttt{pandas} 2.3.3 and \texttt{numpy} 1.23.5 were used throughout.

\subsection{Formal test of a genotype effect}

To test formally whether genotype predicts modulation beyond a visual impression, we fit a cluster-robust logistic regression (session as the cluster unit) predicting FDR-modulated status from a three-level genotype group (control, Sst/Pvalb pooled, Vip). With only eight session-level clusters, standard cluster-robust standard errors are known to be unreliable; Cameron and Miller's practitioner's guide and Cameron, Gelbach, and Miller's original proposal both identify fewer than 20-30 clusters as a regime where the usual asymptotic correction over-rejects, and recommend a wild cluster bootstrap instead~\citep{cameron2015practitioner,cameron2008bootstrap}. We therefore treat the logistic regression as a first pass and additionally ran a wild cluster bootstrap on a linear probability model (OLS) with the same genotype-group design (coefficients sanity-checked to exactly match each group's empirical modulation fraction before testing), using the \texttt{wildboottest} 0.3.2 Python package\footnote{\url{https://github.com/py-econometrics/wildboottest}}. We emphasize that the two p-values compared below come from two different model specifications, a logistic link for the GLM and an identity link for the bootstrapped linear probability model, fit to the identical design matrix and outcome, not two inference procedures applied to one model. This is a deliberate, not incidental, choice: wild-cluster-bootstrap theory is developed primarily for linear estimators, so the linear-probability-model bootstrap is the more defensible of the two tests, and the logistic regression should be read as the first-pass, less rigorous result it was always intended to be. With only eight clusters, the package performed exact full enumeration over all $2^8=256$ possible Rademacher weight assignments rather than Monte Carlo sampling (confirmed directly from the package's internal state, \texttt{full\_enumeration=True}), so the resulting p-values are exact, not subject to simulation noise; with 256 equally likely sign assignments and no ties in the resulting test statistics, the achievable two-sided p-values fall on a grid of multiples of $1/256 \approx 0.0039$, so a reported value such as 0.0625 is exactly $16/256$, not a rounding artifact.

We note a pattern that recurs throughout this paper and state it here as a general principle rather than case by case: every contrast in which a two-session genotype group (Vip-ChR2, the smallest group by cluster count) is involved shows the cluster-robust logistic regression reporting a smaller p-value than the exact wild cluster bootstrap, sometimes by many orders of magnitude, while contrasts not involving this smallest group behave more consistently across methods. This is the expected direction of bias for cluster-robust Wald tests with few clusters, which are known to over-reject, and we treat it as a reason to read logistic p-values as a first-pass heuristic throughout this paper, not only in the specific contrasts where the discrepancy happens to be largest.

\subsection{Restricting to putative direct responders}
\label{sec:direct}

To address the population-versus-targeted-cell distinction raised in the Introduction, we reclassified every FDR-modulated unit (526 of 4{,}112) by its median light-evoked first-spike latency, using only trials in which a spike occurred. There is no universal numeric standard for classifying a unit as directly light-driven rather than network-mediated in the optotagging literature; the Allen Institute's own optotagging documentation states explicitly that "there is no single standard for what metrics to use." We therefore used an explicitly stated, literature-grounded threshold rather than an unstated or arbitrary one: a median light-evoked latency under 10 ms classifies a unit as a putative direct responder, grounded in the approximately 1 ms physiological floor for generating a light-evoked action potential, typical reported ChR2-positive latencies of several milliseconds in other in vivo optotagging studies that use an explicit sub-10 ms criterion, and this paper's own Methods convention that ChR2-driven interneuron responses are typically well under 20 ms. We also computed, but did not use as a second hard gate, the standard deviation of light-evoked latencies (jitter) per unit, since no defensible numeric jitter threshold exists in the literature to cite; jitter is reported only descriptively. We define a unit as a \emph{direct responder} if it is both FDR-modulated and putative-direct, over all 4{,}112 tested units, not only the modulated subset, so that direct-responder fraction is computed on the same denominator as the unrestricted modulation fraction and the two are directly comparable. We reran the same cluster-robust logistic regression and exact wild cluster bootstrap described above on this restricted outcome.

A visual inspection of the 526 FDR-modulated units, decomposed by genotype, suggested a further, separate pattern: that the \emph{proportion} of modulated units classified putative-direct (as opposed to the direct-responder fraction over all units, above) might itself differ by genotype, in a direction opposite the raw modulation-fraction ordering. Because this paper's own stated position is that no descriptive pattern should be reported without the same formal test applied to its headline results, we tested this proportion with the identical cluster-robust logistic regression and exact wild cluster bootstrap used throughout Section~\ref{sec:genotype_test}, restricted to the 526 modulated units with genotype group (control, Sst/Pvalb, Vip-ChR2) as the predictor and putative-direct status as the outcome, session as the cluster unit. We additionally tested, as a direct check on an alternative explanation, whether a unit's FDR evidence strength (its $q$-value) predicts putative-direct classification independent of genotype, using a cluster-robust logistic regression of putative-direct status on $\log_{10}(q\text{-value})$ with and without a genotype term: if weaker, more marginal FDR calls are more likely to be classified direct regardless of genotype, that would indicate the pattern reflects noise contamination of the modulated pool rather than a genotype-specific direct-versus-network balance.

\subsection{Validation}

To confirm the pipeline detects a real effect rather than an artifact, we ran a circular-shift permutation test: for each session, all light-pulse onset times were shifted by the same random offset, wrapped around that session's own spike-time range, and the full pipeline, including pooled FDR correction, was recomputed under this shuffled alignment, repeated 20 times.

\section{Results}

\subsection{Per-session and pooled modulation fraction}

\begin{figure}[htbp]
\centering
\includegraphics[width=0.8\textwidth]{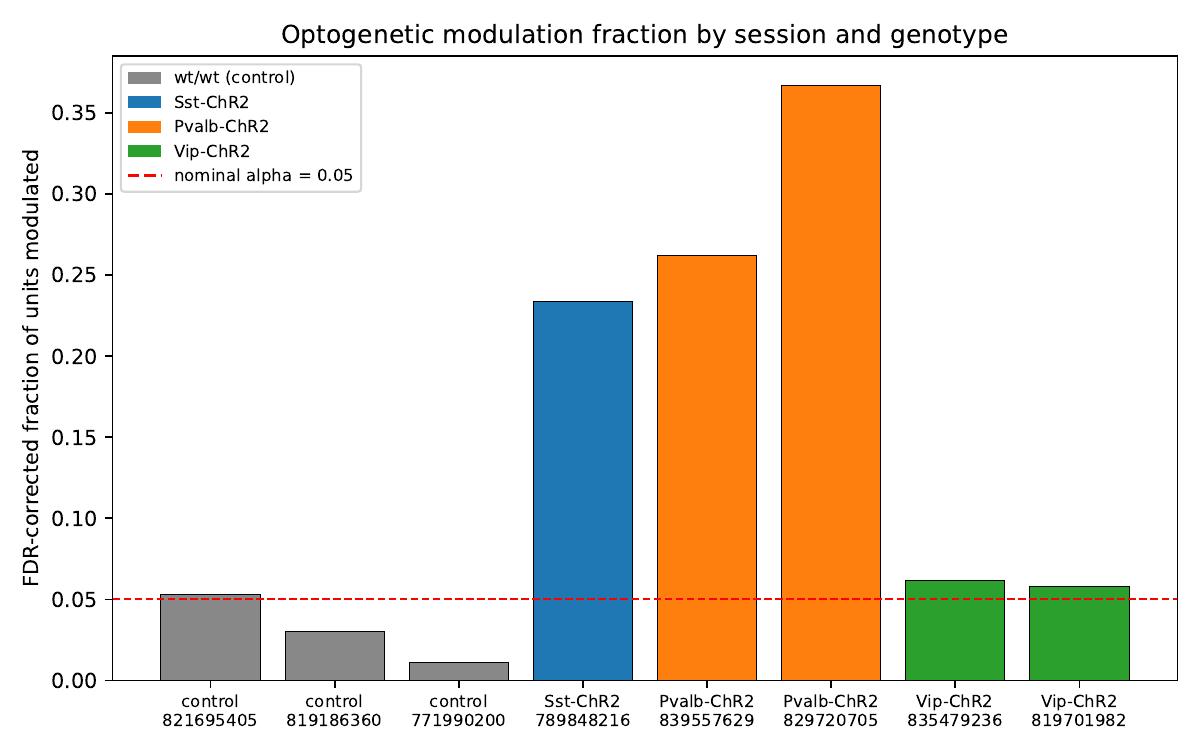}
\caption{FDR-corrected fraction of units modulated by light stimulation, per session, colored by genotype.}
\label{fig:genotype_bar}
\end{figure}

Table~\ref{tab:per_session} and Figure~\ref{fig:genotype_bar} show the per-session and pooled modulation fractions. Sst/Pvalb sessions cluster between 0.23 and 0.37 FDR-corrected; Vip sessions sit consistently near 0.06; wild-type controls fall at or below the nominal 5\% false-positive rate.

\begin{table}[htbp]
\centering
\caption{Fraction of units modulated by light stimulation, per session and pooled, FDR correction applied once across the pooled population.}
\label{tab:per_session}
\begin{tabular}{lrrrr}
\toprule
Session & Genotype & $n$ units & Uncorrected & FDR-corrected \\
\midrule
789848216 & Sst-ChR2 & 415 & 0.3663 & 0.2337 \\
839557629 & Pvalb-ChR2 & 450 & 0.3711 & 0.2622 \\
821695405 & wt/wt (control) & 474 & 0.0992 & 0.0527 \\
829720705 & Pvalb-ChR2 & 529 & 0.4783 & 0.3667 \\
819186360 & wt/wt (control) & 531 & 0.0716 & 0.0301 \\
771990200 & wt/wt (control)\textsuperscript{a} & 546 & 0.0458 & 0.0110 \\
835479236 & Vip-ChR2 & 582 & 0.1581 & 0.0619 \\
819701982 & Vip-ChR2 & 585 & 0.1436 & 0.0581 \\
\midrule
\textbf{Pooled (all 8)} & & \textbf{4{,}112} & \textbf{0.2087} & \textbf{0.1279} \\
\bottomrule
\end{tabular}
\vspace{2mm}
\footnotesize \textsuperscript{a}180 light presentations recorded, versus 300 elsewhere.
\end{table}

\subsection{Formal genotype test: logistic regression versus exact wild cluster bootstrap}
\label{sec:genotype_test}

The cluster-robust logistic regression suggests the per-session separation is statistically reliable: relative to control (model-implied probability 0.0303, matching the empirical mean exactly), Sst/Pvalb has significantly higher odds of modulation (coefficient = 2.5868, z = 6.567, p $<$ 0.001; model probability 0.2934), and Vip has higher odds than control but more modestly (coefficient = 0.7139, z = 2.026, p = 0.043; model probability 0.0600). A direct contrast suggests Sst/Pvalb and Vip differ significantly from each other (log-odds difference = 1.8729, z = 10.419, p $<$ 10$^{-6}$).

Table~\ref{tab:bootstrap} compares these results to the exact wild cluster bootstrap. The linear probability model's coefficients exactly reproduce the same empirical group means (0.0303, 0.2934, 0.0600) as the logistic model, confirming the two specifications describe the identical underlying data before any difference in model type or inference method is considered. Under the bootstrap, Sst/Pvalb's elevation over control remains significant (p = 0.0312), consistent with the logistic result; but the Vip-versus-control contrast (p = 0.125) and the direct Sst/Pvalb-versus-Vip contrast (p = 0.0625) no longer clear the conventional 0.05 threshold. We report this discrepancy rather than the single, more favorable logistic result. We also did not apply a further multiple-comparisons correction across these three wild-bootstrap contrasts; doing so (for example, a Bonferroni threshold of approximately 0.017 for three tests) would make even the surviving Sst/Pvalb-versus-control result only marginal, a further reason to treat the genotype effect as suggestive beyond its single most robust contrast rather than as fully settled.

\begin{table}[htbp]
\centering
\caption{Genotype-effect test: cluster-robust logistic regression versus exact wild cluster bootstrap (8 clusters, 256-way full enumeration) on a linear probability model with the identical design.}
\label{tab:bootstrap}
\begin{tabular}{lrr}
\toprule
Contrast & Logistic GLM (cluster-robust) & Wild cluster bootstrap (exact) \\
\midrule
Sst/Pvalb vs. control & p $<$ 0.001 & p = 0.0312 \\
Vip vs. control & p = 0.043 & p = 0.125 \\
Sst/Pvalb vs. Vip (direct) & p $<$ 10$^{-6}$ & p = 0.0625 \\
\bottomrule
\end{tabular}
\end{table}

The pooled Sst/Pvalb category is itself worth decomposing rather than treating as a single unit: it contains only one Sst session against two Pvalb sessions, so the pooled contrast is necessarily weighted toward Pvalb. Examined individually, the single Sst session (0.2337 FDR-corrected) and the two Pvalb sessions (0.2622 and 0.3667) occupy a similar, overlapping range (Table~\ref{tab:per_session}), which is reassuring, but with only one Sst session we cannot rule out that the pooled effect is disproportionately driven by Pvalb specifically, nor can we say whether Sst and Pvalb would remain indistinguishable from each other with more sessions of each.

The honest conclusion is narrower than the logistic model alone would suggest: the raw, descriptive separation between Sst/Pvalb and Vip/control sessions is large and consistent, and Sst/Pvalb's elevation above a true no-opsin baseline is robust to the choice of inference method; but formally distinguishing Vip from control, Sst/Pvalb from Vip, and Sst from Pvalb, with only eight session-level clusters, is underpowered by the more appropriate small-cluster method. More sessions, particularly more Vip, Sst, and control sessions, are needed before these contrasts should be treated as confirmed rather than suggestive.

\subsection{Restricting to putative direct responders: a robust effect confirmed, an ambiguity resolved, and an apparent reversal retracted}
\label{sec:direct_results}

Of the 526 FDR-modulated units, 152 (28.9\%) met the putative-direct criterion (median evoked latency under 10 ms) and 374 (71.1\%) did not. Table~\ref{tab:direct_fraction} reports the restricted direct-responder fraction by session, computed over all 4{,}112 units (not only the modulated subset), so it is directly comparable to Table~\ref{tab:per_session}. Every group's fraction drops substantially relative to the unrestricted metric, as expected, since network-mediated responses are now excluded: pooled, control falls from 0.0303 to 0.0129, Sst/Pvalb from 0.2934 to 0.0775, and Vip from 0.0600 to 0.0206. The ordering (control $<$ Vip $<$ Sst/Pvalb) is preserved, but the gap between Vip and control narrows considerably in relative terms.

\begin{table}[htbp]
\centering
\caption{Direct-responder fraction (FDR-modulated AND putative-direct, median evoked latency $<$10 ms) by session, computed over all tested units in that session.}
\label{tab:direct_fraction}
\begin{tabular}{lrrr}
\toprule
Session & Genotype & $n$ units & Direct-responder fraction \\
\midrule
789848216 & Sst-ChR2 & 415 & 0.0482 \\
839557629 & Pvalb-ChR2 & 450 & 0.1000 \\
821695405 & wt/wt (control) & 474 & 0.0401 \\
829720705 & Pvalb-ChR2 & 529 & 0.0813 \\
819186360 & wt/wt (control) & 531 & 0.0019 \\
771990200 & wt/wt (control) & 546 & 0.0000 \\
835479236 & Vip-ChR2 & 582 & 0.0120 \\
819701982 & Vip-ChR2 & 585 & 0.0291 \\
\midrule
\textbf{Pooled by group} & control & 1{,}551 & \textbf{0.0129} \\
 & Sst/Pvalb & 1{,}394 & \textbf{0.0775} \\
 & Vip & 1{,}167 & \textbf{0.0206} \\
\bottomrule
\end{tabular}
\end{table}

The formal genotype test on this restricted outcome resolves one ambiguity from Section~\ref{sec:genotype_test} and leaves another unchanged. Table~\ref{tab:direct_bootstrap} shows that Sst/Pvalb's elevation over control remains significant under both the logistic GLM (p = 0.033) and the exact wild cluster bootstrap (p $<$ 0.0001), if anything more decisively than under the unrestricted metric. Vip-versus-control, previously ambiguous (logistic p = 0.043 versus bootstrap p = 0.125, disagreeing with each other), is now clearly non-significant by both methods (logistic p = 0.604, bootstrap p = 0.563): restricting to direct responses removes most of Vip's apparent elevation over control, consistent with that elevation having been predominantly network-mediated. The direct Sst/Pvalb-versus-Vip contrast, however, is not resolved: the logistic GLM remains significant (p = 1.3$\times10^{-4}$) while the exact wild cluster bootstrap gives p = 0.0625, numerically identical to the unrestricted analysis. We do not interpret this exact numerical match as a meaningful equivalence; with only eight clusters the bootstrap's achievable p-values form a coarse, discrete grid, and an identical value across two different outcome definitions may simply reflect that grid rather than a substantive finding.

\begin{table}[htbp]
\centering
\caption{Genotype-effect test on the restricted direct-responder outcome.}
\label{tab:direct_bootstrap}
\begin{tabular}{lrr}
\toprule
Contrast & Logistic GLM (cluster-robust) & Wild cluster bootstrap (exact) \\
\midrule
Sst/Pvalb vs. control & p = 0.033 & p $<$ 0.0001 \\
Vip vs. control & p = 0.604 & p = 0.563 \\
Sst/Pvalb vs. Vip (direct) & p = 1.3$\times10^{-4}$ & p = 0.0625 \\
\bottomrule
\end{tabular}
\end{table}

Decomposing the 526 modulated units by genotype initially suggested a striking pattern: the proportion classified as putative direct responses was 20.6\% for Sst-ChR2, 28.2\% for Pvalb-ChR2, and 34.3\% for Vip-ChR2, the opposite ordering from the raw modulation-fraction results in Table~\ref{tab:per_session}. On its own, this descriptive pattern would read as a genuine finding: Sst/Pvalb sessions have a larger raw modulation fraction, but, by this reading, a larger share of that modulation would be network-mediated rather than direct, relative to Vip. We tested this proportion formally, using the same cluster-robust logistic regression and exact wild cluster bootstrap applied to every other genotype contrast in this paper (Methods, Section~\ref{sec:direct}), with the three-level genotype grouping (control, Sst/Pvalb, Vip-ChR2) and session as the cluster unit. Table~\ref{tab:reversal_test} reports the result: none of the three pairwise contrasts approaches significance under the exact bootstrap (p = 0.75-0.84 for all three), and the pooled, genotype-level proportions tell a different story once control is included on equal footing: control (42.6\%) is actually \emph{higher} than Vip (34.3\%), which is higher than Sst/Pvalb (26.4\%). Because true no-opsin control sessions contain no direct ChR2-driven responses by construction, a pattern in which control exceeds every ChR2-expressing group cannot be a genotype-dependent direct-versus-network effect; it is evidence that the apparent reversal is an artifact of the classification criterion's limited specificity (discussed next) compounded by small, noisy per-session counts, not a biological pattern that happened to also appear in control.

\begin{table}[htbp]
\centering
\caption{Formal test of the direct-response proportion among modulated units only (526 units), by genotype, using the same cluster-robust logistic regression and exact wild cluster bootstrap (8 clusters, full enumeration) as Tables~\ref{tab:bootstrap} and~\ref{tab:direct_bootstrap}. Pooled proportions include control for comparison; a true no-opsin group cannot show a genuine direct ChR2 response by construction.}
\label{tab:reversal_test}
\begin{tabular}{lrr}
\toprule
Contrast & Pooled proportions & Wild cluster bootstrap (exact) \\
\midrule
Sst/Pvalb vs. control & 0.264 vs. 0.426 & p = 0.8438 \\
Vip vs. control & 0.343 vs. 0.426 & p = 0.8125 \\
Sst/Pvalb vs. Vip & 0.264 vs. 0.343 & p = 0.7500 \\
\bottomrule
\end{tabular}
\end{table}

The underlying per-session counts explain why: direct-response proportion among modulated units varies enormously from session to session, by far more than any apparent between-genotype gap, and this variance is present \emph{within} the control group alone. The three control sessions' direct-among-modulated proportions are 0\% (0/6 modulated units), 6.25\% (1/16), and 76.0\% (19/25); the three Sst/Pvalb sessions range from 20.6\% to 38.1\%; the two Vip sessions range from 19.4\% to 50.0\%. A single control session (76.0\%) exceeds every other session of any genotype, driven by a small denominator (25 modulated units) in which a modest absolute count of spuriously fast-latency units dominates the proportion. We report this as a cautionary result, not a confirmatory one: an apparent reversal that, reported as a bare descriptive percentage, would have been a compelling secondary finding did not survive the identical small-cluster-robust test that confirmed this paper's actual central result (Table~\ref{tab:bootstrap}), and we think the contrast between how convincing the raw percentages look and how clearly the formal test rejects them is itself an argument for applying this kind of test to secondary, descriptive patterns and not only to a paper's headline claim.

This restriction also surfaced a limitation of the classification criterion itself, which both explains the pattern above and is worth reporting on its own terms: 20 of 47 (42.6\%) FDR-modulated units in the three true no-opsin wild-type control sessions also met the under-10-ms putative-direct criterion. Since these sessions express no ChR2, any unit so classified cannot reflect a genuine direct photocurrent response; the 10 ms threshold alone does not fully separate true fast optogenetic responses from fast-onset background activity or light-associated artifact (for example, residual visual responses to the stimulation light itself). This means the direct-responder fraction, while a meaningful refinement over the unrestricted metric for the paper's central, robust contrast (Sst/Pvalb versus control), should be read as reducing but not eliminating the population-versus-targeted-cell ambiguity, and the pooled control direct-responder fraction (0.0129 of all tested units, equivalently 42.6\% of control's own modulated units) is better understood as a noise floor for this specific classification rule than as a true estimate of zero.

As a direct check on a specific alternative explanation for the initial descriptive pattern, raised during review, that Vip's modulated pool might simply contain more marginal, lower-confidence FDR calls than Sst/Pvalb's, and that such marginal calls might be more prone to a spuriously early first spike regardless of genotype, we compared each modulated unit's FDR $q$-value (a continuous measure of evidence strength, already computed for every unit) across genotype groups and tested whether $q$-value predicts putative-direct classification independent of genotype. Sst/Pvalb's modulated units have systematically stronger FDR evidence (mean $q = 0.0064$, median $q < 0.0001$) than Vip's (mean $q = 0.0102$, median $q = 0.0023$; Mann-Whitney p = 0.0217), consistent with Vip's modulated pool containing more borderline calls. A cluster-robust logistic regression of putative-direct status on $\log_{10}(q\text{-value})$ alone (pooled across genotypes) gives a coefficient in the direction predicted by noise contamination, weaker evidence associated with a higher chance of being classified direct, but it does not reach significance at this sample size (coefficient $=-0.099$, p = 0.112); adding genotype to the same model leaves neither the genotype terms nor the $q$-value term significant (all p $>$ 0.10). We read this as consistent with, but not strong independent confirmation of, the noise-contamination explanation: the data point in the predicted direction without being decisive, which is itself in keeping with the formal test above showing the underlying pattern is not statistically distinguishable from noise in the first place.

As a further, more limited check on whether the original descriptive pattern could instead be a density or sampling artifact, we compared mean recorded units per session and the fraction of units in cortical visual-area structures across genotype groups. Vip sessions had, if anything, more recorded units on average (583.5) than Sst/Pvalb (464.7) or control (517.0) sessions, and the cortical-structure fraction was similar across groups (0.515, 0.470, and 0.518, respectively). This is a coarse, observational check using only the \texttt{ecephys\_structure\_acronym} field already in our per-unit output, not a full analysis, and it cannot rule out finer-grained local density differences that this field cannot resolve, but it does not support the simplest version of a sparser-Vip-sampling explanation, and in any case the formal test above already shows there is no genotype-level pattern left to explain.

\subsection{Validation and a composition-dependent artifact}

The real uncorrected fraction (0.2087) is 6.8 times the shuffled-null mean (0.0307, SD 0.0033) from the circular-shift permutation test, and after FDR correction the shuffled null is exactly zero across all 20 shuffles while the real result (0.1279) remains clearly nonzero. This shuffle count is modest by design: each shuffle reruns the complete per-unit Cox and log-rank pipeline across all 4{,}112 units and all eight sessions' underlying recordings, which is computationally expensive; 20 shuffles are enough to show the real result is a clear outlier relative to the null distribution's mean and standard deviation, exactly the claim we make, but not enough for a precise p-value on the pooled fraction itself, which we do not claim either.

A smaller, earlier three-session version of this analysis (one Sst-ChR2, one Pvalb-ChR2, one wild-type control session) gave a pooled FDR-corrected estimate of 0.1934, noticeably higher than the eight-session estimate of 0.1279. This difference is not evidence that the underlying biology changed; it reflects that the three-session batch happened to include two ChR2-expressing sessions against only one control, while the eight-session pool is more balanced. The pooled estimate is therefore sensitive to which sessions happen to be included, which is itself an argument for reporting modulation fraction by cell type rather than as a single pooled number, and for treating any pooled estimate from a small number of sessions as provisional until checked against a larger, more balanced sample.

\section{Discussion}

The central, robust finding is that ChR2 expression itself substantially elevates modulation fraction over true no-opsin control, under both the unrestricted population-level metric and the restricted direct-responder metric, and under both a standard cluster-robust test and an exact small-cluster-appropriate bootstrap. This opsin-presence effect is the paper's most secure result. The cell-type question beneath it, whether Sst/Pvalb and Vip interneurons differ from each other in response rate, is less settled, and we were able to resolve one part of it and had to retract another. On the unrestricted, population-level metric, Vip sessions show a lower raw modulation fraction than Sst/Pvalb, and restricting to direct responders resolves Vip-versus-control toward clear non-significance by both inference methods, a genuine narrowing of ambiguity. A second pattern, that among modulated units a larger proportion of Vip's response was classified as direct rather than network-mediated, the opposite ordering from the raw fraction, looked at first like a second finding worth a biological explanation. It is not: formally tested with the same cluster-robust bootstrap used for every other genotype contrast in this paper, none of the three pairwise contrasts on this proportion approach significance, and the pooled, true no-opsin control group has the highest proportion of the three, which is impossible to explain biologically and points instead to classification noise and small per-session counts (Section~\ref{sec:direct_results}). We do not offer a biological interpretation of a pattern that did not survive its own formal test; the correct lesson is methodological, not neurobiological, and we draw it out below.

This dataset also cannot fully rule out a measurement-level explanation for part of the Sst/Pvalb-versus-Vip gap that \emph{does} remain unresolved (the direct Sst/Pvalb-versus-Vip contrast in Table~\ref{tab:direct_bootstrap}, bootstrap p = 0.0625 both before and after restriction): a cell-type-dependent difference in ChR2 expression level under the Ai32 reporter construct. Leaky and off-target expression of ChR2 in Ai32 mouse lines is itself independently documented~\citep{prabhakar2019leaky}, and some units scored as non-modulated, or as network-mediated rather than direct, could in principle reflect expression differences rather than a genuine cell-type difference in intrinsic response properties; we had no independent histological or photocurrent verification of expression for individual recorded units. We also found that our own direct-responder criterion has limited specificity, misclassifying 42.6\% of modulated units in true no-opsin controls as putative-direct, so the direct-versus-network distinction itself should be read as a noisy refinement, not a clean separation; this same limited specificity, combined with small per-session modulated-unit counts, is what we conclude produced the retracted reversal pattern discussed above, rather than any cell-type-dependent direct-versus-network balance.

Four self-identified corrections run through this reanalysis, and we see them, collectively, as being as important as any single numerical result that survived. First, a smaller, earlier three-session version of this analysis produced a pooled estimate that was an artifact of session composition, not biology, resolved only once more sessions were added. Second, a standard cluster-robust test suggested all three genotype contrasts on the unrestricted metric were significant, and only an exact small-cluster-appropriate bootstrap revealed that two of the three do not survive. Third, restricting to putative direct responders, motivated by a reviewer's correct observation that the unrestricted metric conflates targeted and network effects, resolved the Vip-versus-control ambiguity toward non-significance. Fourth, a descriptive pattern in the direct-response proportion among modulated units, which an earlier draft of this paper reported as a genuine and unexpected reversal, did not survive the same formal test applied to every other claim in this paper, and we retract it here rather than let a compelling-looking percentage stand without the scrutiny the rest of the paper insists on. None of these four corrections was forced by the statistical framework itself; each came from checking an intermediate result against an independent fact, a larger sample, a more appropriate small-sample method, or the same formal test already used elsewhere, rather than trusting a result because the pipeline ran without error or because a percentage looked compelling on its own. We would rather report a paper whose headline contrasts shift, and whose secondary claims are sometimes withdrawn entirely, under harder checks than one that only reports the analyses that happened to look cleanest first.

\section{Limitations}

The single fixed 50 ms post-pulse window, applied uniformly to pulses, trains, and ramps, is a simplification; for the second-long trains and ramps it characterizes only the initial response window, not the full stimulus duration. Session selection was governed entirely by unit count, not by a designed, balanced sampling of Cre-lines; the resulting genotype mix (three Pvalb, three wild-type, two Vip, one Sst session) is informative but not representative, and 18 of the 26 available \texttt{functional\_connectivity} sessions, including other Cre-lines not represented here, remain unanalyzed. The genotype-effect test itself has only eight session-level clusters (two of them Vip, only one of them Sst), and as shown directly above, this is not a hypothetical concern: the exact wild cluster bootstrap overturns the cluster-robust logistic regression's two weaker contrasts on the unrestricted metric, leaving only Sst/Pvalb's elevation over control as robust to the choice of method on both outcome definitions.

The direct-responder restriction introduced in Section~\ref{sec:direct} addresses the population-versus-targeted-cell ambiguity in the unrestricted metric, but introduces its own, explicitly reported limitation: our 10 ms latency criterion, grounded in the literature but necessarily a single stated choice among several defensible options, misclassifies 42.6\% of modulated units in true no-opsin control sessions as putative-direct, so it reduces rather than eliminates the conflation it is meant to address. This same limited specificity is, by our own formal testing, the most likely explanation for an initially striking direct-versus-network pattern (Vip's modulated units appearing proportionally more often direct than Sst/Pvalb's) that we report in Section~\ref{sec:direct_results} as retracted rather than confirmed: a cluster-robust bootstrap test of this specific proportion found no significant genotype contrast (p = 0.75-0.84), and the pooled true no-opsin control group had the highest proportion of the three, which cannot reflect a real direct-response difference since control units express no opsin. We flag this explicitly as a case where a compelling descriptive pattern did not survive the same formal test applied to this paper's confirmed results, and as a caution against reporting any genotype-stratified proportion, however striking, without that test. The Sst/Pvalb-versus-Vip contrast on the (confirmed, not retracted) direct-responder fraction itself remains unresolved by the exact wild cluster bootstrap under both the unrestricted and restricted outcome, at an identical p-value we explicitly do not interpret as a meaningful coincidence given the coarseness of an 8-cluster bootstrap's achievable p-value grid (multiples of $1/256$). We also cannot rule out opsin-expression failure, or a cell-type-dependent difference in expression level, as an alternative explanation for this unresolved contrast, as noted above. Our structure-based check against a density or sampling artifact is coarse and observational, not a full analysis, and cannot rule out finer-grained local confounds, though it is moot for the retracted reversal specifically since no genotype-level pattern remains there to explain. Across the six genotype-comparison tests reported in this paper (three contrasts, two outcome definitions), we did not apply a further family-wise correction beyond the per-metric Bonferroni note in Section~\ref{sec:genotype_test}; a family-wise threshold across all six (approximately 0.0083) would still be cleared by the restricted Sst/Pvalb-versus-control result (bootstrap p $<$ 0.0001) but by no other contrast, which is consistent with our framing of only that one contrast as fully robust. Finally, this analysis is confined to one dataset family (Allen Institute Visual Coding Neuropixels) and one opsin (ChR2); whether the patterns observed here generalize to other preparations, opsins, or stimulation parameters is untested.

\section{Conclusion}

Using a censoring-aware survival framing, a single pooled multiple-comparisons correction, a formal small-cluster-appropriate statistical test, and a restriction to putative direct responders that distinguishes targeted-cell effects from network-mediated ones, we find that ChR2 expression robustly and substantially elevates optogenetic modulation fraction over true no-opsin control, under every method and outcome definition we tried. The finer question of cell-type dependence within ChR2-expressing interneurons is only partly resolved: Vip-versus-control becomes clearly non-significant once restricted to direct responders, while Sst/Pvalb-versus-Vip remains unresolved under both outcome definitions. A third pattern, an apparent reversal in which Vip's modulated units looked proportionally more often direct than Sst/Pvalb's, did not survive formal testing with the same small-cluster-robust bootstrap used throughout this paper, and we report its retraction, not its confirmation, as part of this paper's result. We also show that a smaller, less balanced session pool can produce a misleading single-number estimate, that an unrestricted, population-level modulation metric can obscure a distinction, direct versus network-mediated response, that matters for interpretation, and that a descriptive, genotype-stratified proportion can look like a finding and not be one until the same formal test applied to a paper's headline claim is applied to it as well. Together, these results support reporting optogenetic modulation fraction by cell type and by direct-versus-network status rather than as a single aggregate figure, support a small-cluster-robust statistical test as a standard check whenever a genotype or condition effect is assessed across a handful of sessions, whether that effect is a paper's central claim or a secondary descriptive pattern, and argue for survival-based, FDR-corrected, bootstrap-validated, latency-stratified quantification as a reusable addition to how circuit-dissection studies characterize stimulation efficacy.

\section*{Data and Code Availability}
All analysis code, including session-selection logic, the survival-model pipeline, the permutation validation script, the genotype-effect regression and wild cluster bootstrap, and per-unit result tables, is available at \url{https://github.com/SID-6921/optogenetics-research} (private repository; access available on request). The underlying Neuropixels recordings are publicly available from the Allen Institute via the AllenSDK \texttt{EcephysProjectCache} interface and require no special access.

\section*{Ethics Declaration}
This work is a secondary computational reanalysis of previously collected, publicly released data. All animal procedures associated with the original data collection were approved under the Allen Institute's own Institutional Animal Care and Use Committee protocols; no new animal or human subjects procedures were performed for this work.

\section*{Funding}
This work was conducted independently and received no dedicated funding.

\section*{Conflict of Interest}
The author declares no conflict of interest.

\section*{Author Contributions}
S.N.: Conceptualization, Methodology, Software, Formal Analysis, Validation, Writing - Original Draft, Writing - Review \& Editing.

\bibliographystyle{vancouver}

\end{document}